\documentstyle[12pt]{article}
\begin{document}
\baselineskip 22pt%+2pt

\hspace{11cm} TECHNION -PH-92-1

\hspace{11cm} {\em January 1992}

\vspace{1cm}

\begin{center}
{\bf String Field Theory for $d \leq 0$ Matrix Models via
Marinari-Parisi} \\
\vspace{0.50cm}

Joshua Feinberg\footnote{Supported in part by the Fund for Basic
Research administered by the Israel Academy of Sciences and
Humanities}$^,$\footnote{e-mail address: PRH74JF@TECHNION.BITNET}\\

Department of Physics, Technion - Israel Institute of Techology\\
Haifa, 32000, Israel.
\vspace{0.50cm}

\noindent{\bf ABSTRACT}
\end{center}
\vspace{0.25cm}

We generalize the Marinari-Parisi definition for pure two dimensional
quantum gravity $(k = 2)$ to all non unitary minimal multicritical
points $(k \geq 3)$.  The resulting interacting Fermi gas theory is
treated in the collective
field framework.  Making use of the fact that the matrices evolve in
Langevin time, the Jacobian from matrix coordinates to collective modes
is similar to the corresponding Jacobian in $d = 1$ matrix models.
The collective field theory is analyzed in the planar limit.  The saddle
point eigenvalue distribution is the one that defines the original
multicritical point and therefore exhibits the appropriate scaling
behaviour.
Some comments on the nonperturbative properties of the collective field
theory as well as comments on the Virasoro constraints associated with
the loop equations are made at the end of this letter.  There we also
make some remarks on the fermionic formulation of the model and its
integrability, as a nonlocal version of the non linear Schr\"{o}dinger
model.

%ex\vspace{2cm}
%January 1992
\pagebreak

Hermitian one matrix models defined at a single point\footnote
{i.e. - matrix models defined over a zero dimensional space time.  We
will refer to them as zero dimensional models.  The stabilized models,
in
which matrices depend upon the Langevin time coordinate will be referred
to as the one dimensional models.   One should not, of course, mix
this
terminology with ``dimension'' counting in target space - i.e. values of
the central charge of matter coupled to gravity.}
describe at their double scaling multicritical points minimal nonunitary
conformal matter coupled to two dimensional quantum gravity \cite{GR MG}
\cite{DG SH} \cite{BR KZ ZA}.

Pure gravity $(k = 2)$ as well as all higher multicritical points of
even order $k \ (k = 4, 6, ...)$ exhibit non perturbative ambiguities,
instabilities and violations of the Schwinger-Dyson (loop) equations
\cite{DG SH SEI}   \cite{SH} \cite{ZJ GI} \cite{DA1}
\cite{DA2}\cite{DA3}.

These  problems can be traced back to the fact that the critical matrix
potentials of even order $k$ are bottomless.  Thus a sensible
definition of these models should correspond to well-defined stabilized
matrix potentials.

Bottomless matrix potentials occur also in certain multimatrix models
describing unitary matter coupled to two dimensional quantum gravity
such as the two matrix model corresponding to the Ising case
\cite{ISING}.
Thus, the problem of stabilization is associated not only with one
matrix models.

Marinari and Parisi \cite{PM(1)} have suggested a possible way out of
this
difficulty in the case of pure gravity by supersymmetrizing the model.

The bosonic sector hamiltonian in \cite{PM(1)} is also the forward
Fokker-Planck hamiltonian associated with the Langevin equation whose
force term is minus the gradient of the original zero dimensional matrix
action.

Therefore, the definition of pure two dimensional quantum gravity in
\cite{PM(1)} is equivalent to the stabilization procedure developed in
\cite{GR HAL} for bottomless actions, as far  as the bosonic sector of
the former is concerned.

The stabilized pure gravity model of \cite{PM(1)} was further analyzed
by \cite{KAR MIG}, where the one eigenvalue double scaled hamiltonian
was
extracted, and a nonperturbative ambiguity free analysis of the density
of particles and density of state was made.

The forward Fokker-Planck hamiltonian\footnote{In \cite{BA DO
SHE} the
existence of such an ${\cal H}$ was postulated, without specifying
its
details to define ``time ordered''correlators in the zero dimensional
theory}, used in \cite{PM(1)} \cite{KAR MIG} for the pure gravity case,
reads for a general matrix potential $U(\Phi)$
\begin{eqnarray}
{\cal H} = \frac{1}{2} Tr [(-\frac{\partial}{\partial \Phi} +     %1
\frac{\beta}{2} U^\prime(\Phi))(\frac{\partial}{\partial \Phi} +
\frac{\beta}{2} U^\prime (\Phi))]
\end{eqnarray}
Here $\Phi$ is an hermitian matrix, depending on the Langevin time
coordinate (i.e. the bosonic coordinate of superspace used in
\cite{PM(1)}).

${\cal H}$ in eq. (1) is a well defined Schr\"{o}dinger operator.  Its
potential is clearly bounded from below and grows to plus
infinity as the matrix eigenvalues become infinite.

Therefore, a well defined unique normalizable ground state vector
$\Psi_0(\Phi)$ exists.  Indeed, if $U(\Phi)$ is bounded from below such
that the Boltzman weight of the zero dimensional matrix model is
normalizable this ground state is given by
\begin{eqnarray}
\Psi_0(\Phi) = \frac{1}{\sqrt{{\cal {Z}}}}e^{\frac{-\beta}{2}T_rU(\Phi)}
\
, \ \ {\cal Z} = \int d^{N^2} \Phi e^{- \beta Tr U(\Phi)}            %2
\end{eqnarray}

In this case the vacuum energy is strictly zero and supersymmetry
is not broken. Moreover, expectation values of
operators, all at infinite Langevin time project only onto the
ground state $\Psi_0$, and are identical to the
corresponding correlators in the original zero dimensional matrix model:
\begin{eqnarray}
< \Psi_0 \mid {\cal O}_1(\Phi) \cdots {\cal O}_n(\Phi)\mid \Psi_0 >
%3
= \frac{1}{{\cal Z}} \int d^{N^2} \Phi e^{-\beta Tr
U(\Phi)}{\cal O}_1(\Phi) \cdots {\cal O}_n(\Phi)
\end{eqnarray}

If $U(\Phi)$ is unbounded from below, the zero dimensional Boltzman
weight is unnormalizable and the corresponding matrix model exists only
at a saddle point level.  Supersymmetry is broken and the vacuum energy
$E_0$ is positive.  Alternatively - the appropriate Langevin equation
has only runaway solutions and the Fokker-Planck probability density at
any finite portion of matrix eigenvalue space, decays  asymptotically
in Langevin time $t$ as $e^{-E_0t}$ \cite{P} \cite{ZJ}\footnote{For
this property
to hold also in the double scaling limit, we must ensure that the vacuum
remains nondegenerate even as $N \rightarrow \infty$, i.e. - that the
energy eigenvalue $E_1$ of the first excited state of ${\cal H}$ does
not coalesce with $E_0$. As was shown in \cite{PM(1)} the mass gap
$E_1-E_0$
double scales in the WKB approximation for $k = 2$.   Moreover, we will
show that it double scales for any value of $k$.  Thus the vacuum state
remains nondegenerate.  This was tacitly assumed in \cite{GR HAL}.}.

However, Fokker-Planck averages of operators normalized by the
Fokker-Planck average of the unit operator are well defined as the
Langevin time goes to infinity, and correspond to
\begin{eqnarray}
<{\cal O}_1 (\Phi) \cdots {\cal O}_n (\Phi) >_{t \rightarrow \infty} =
%4
\int d^{N^2} \Phi \mid \Psi_0 (\Phi) \mid^2 \ {\cal O}_1(\Phi) \cdots
{\cal O}_n(\Phi)_{\mid_{t = \infty}}
\end{eqnarray}
Here $\Psi_0(\Phi)$ is the normalizable ground state of ${\cal H}$, and
all the operators on the r.h.s. of Eq. (4) are at $t = \infty$.  We
thus consider Eq. (4) as
the stabilized definition for correlators in case of bottomless matrix
potentials.

As is well known, the laplacian over hermitean matrices acquires the
form
\begin{eqnarray}
- Tr \frac{\partial^2}{\partial \Phi^2} = - \frac{1}{\Delta(x)} \
\sum^N_{i=1} \ \frac{\partial^2}{\partial x^2_i} \ \Delta(x) +
\left( \matrix{\mbox{nonsinglet} \  U(N) \ \mbox{angular} \cr
\mbox{momentum terms} } \right)
\end{eqnarray}
where $x_i$ are the matrix eigenvalues and $\Delta(x_i)$ is the
Vandermonde determinant.  This leads to the mapping of eigenvalue
dynamics onto that of a one dimensional Fermi gas \cite{BP IZ}.
Clearly,
the ground state $\Psi_0(\Phi)$ mentioned above is a $U(N)$ singlet.

For a generic potential $U(\Phi)$, the hamiltonian in Eq. (1) contains
a two body interaction term among eigenvalues\footnote{Note that for
$U^\prime(\Phi) = \sum_{0 \leq n} C_n Tr \Phi^n$ we obtain ${\cal
H}_{int} = \sum_{0 \leq n} C_n \sum^{n - 1}_{\ell = 0} Tr \Phi^{\ell} Tr
\Phi^{n-1-\ell},$  therefore from the point of view of the (noncritical)
one dimensional matrix theory ${\cal H}_0 = \frac{1}{2} Tr ( -
\frac{\partial^2}{\partial \Phi^2} + \frac{\beta^2}{4}
U^\prime(\Phi)^2)$, whose eigenvalues form a noninteracting Fermi gas in
the singlet sector,
${\cal H}_{int}$ may be interpreted as higher curvature terms
\cite{HIGHER CURVATURE} that push the system to its multicritical point.}
\begin{eqnarray}
{\cal H}_{int} = - \frac{\beta}{4} \ Tr \ \frac{\partial}{\partial \Phi}
\ U^\prime(\Phi) =
- \frac{\beta}{4} \sum_{i,j} \frac{U^\prime(x_i) - U^\prime(x_j)}{x_i -
x_j}                                                               %6
\end{eqnarray}
Thus, generally -- the one dimensional gas of eigenvalues is an
interacting Fermi gas.

The matrix potential of minimal degree that leads to the $k=2$
multicritical (pure gravity) point is
$U_2(\Phi) = - \frac{\lambda}{6} \Phi^3 + \frac{1}{2} \Phi^2 + \Phi$
where we have followed the normalizations of \cite{NEUB}.
The critical coupling constant corresponding to the $k=2$ point is
$\lambda_c = 1$.  The matrix potential $U_2(\Phi)$ is bottomless.
Therefore the stabilized theory will exhibit spontaneous
supersymmetry breaking. This issue was analyzed in
\cite{AT DAB}.
In this case, the interaction term in Eq. (6) reduces to an interaction
of the eigenvalues with a constant background field, proportional to the
number of eigenvalues, N, namely, \
${\cal H}_{int} =   \frac{\beta N}{4} T_r (\lambda \Phi - {\bf 1}).$

Therefore, the singlet sector of Eq. (1) reduces effectively to a system
of $N$ non interacting Fermions in one dimension\footnote{In
\cite{PM(1)} $N/\beta$ was set to 1 and $\lambda$ was varied, while in
\cite{KAR MIG}
$\lambda$ was set to its critical value $\lambda_c = 1$ and $N/\beta$
was varied around its critical value $(N/\beta)_c = 1$.} \cite{PM(1)}
\cite{KAR MIG}
\begin{eqnarray}
{\cal H}_{singlet} =                                    %9
\beta^2 \ \sum^N_{i=1} ( - \frac{1}{2 \beta^2} \
\frac{\partial^2}{\partial x^2_i} + V(x_i)) =
\beta^2 \sum^N_{i=1} h(x_i, p_i)
\end{eqnarray}
where the external one body potential is
\begin{eqnarray}
V(x) = \frac{1}{8} (U^\prime_2 (x))^2 + \frac{1}{4} \        %10
\frac{N}{\beta}(\lambda x-1)
\end{eqnarray}

The ground state of the Fermi gas described in Eqs. (7) and (8) is
obtained by filling the first $N$ one particle levels of $V(x)$.  The
associated Fermi energy must be evaluated self-consistently from the
N-dependent potential $V(x)$.  Therefore, unlike the case of the d=1
model, the Fermi energy is not a free parameter that can be used to
define the double scaling limit.

In order to study the higher stabilized multicritical points $(k \geq
3)$, one has to cope with the interaction term in Eq. (6).  Since only
the U(N)-singlet ground state $\Psi_0(\Phi)$ of ${\cal H}$ is involved
-- it is natural to analyze the interacting gas in terms of the Fermion
density operator -- i.e., the collective field $\phi(x)$ associated with
the matrix $\Phi$.

Following \cite{COLL} \cite{DAS JEV} we define
\footnote{We have used
the normalization of \cite{GR KL1}.}
the collective field as
\begin{eqnarray}
\phi(x) = \frac{1}{\beta} \ T_r \delta(1 \cdot x - \Phi) =
\frac{1}{\beta} \ \sum^N_{i=1} \ \delta(x-x_i)              %13
\end{eqnarray}
which implies that $\phi(x)$ is a non negative operator and obeys the
normalization condition
\begin{eqnarray}
\int \phi(x) dx = \frac{N}{\beta}
\end{eqnarray}

Since $\Phi$ depends on the Langevin time $t$ and its dynamics is fixed
by ${\cal H}$ in Eq. (1), the Jacobian of the transformation from the
matrix eigenvalue variables to the collective field is exactly the same
Jacobian as in the $d = 1$ case \cite{DAS JEV}
\footnote{
The use of precisely this Jacobian is
dictated by the very definition of the supersymmetrized model in
accordance with \cite{CO DE} \cite{JEV 777}, bypassing the need to
invoke arguments
of the type used by \cite{CO DE}, or postulates about the form of the
zero
dimensional partition function as in \cite{JEV 777}, that are needed if
one makes the transformation to collective modes directly in the zero
dimensional matrix model.}

In this letter, we concentrate on the planar approximation to ${\cal
H}$ in order to establish the
fact that the $k-th$ order multicritical behaviour is respected by our
formalism.

Discussing the exact loop equations and the associated non perturbative
effects is deferred to a subsequent publication \cite{JF}.

The planar collective field action $S_0[\phi] = \beta^3 \int {\cal
L}_{(0)}dxdt$ \ for
the matrix  Hamiltonian ${\cal H}$ in Eq. (1) is given by \cite{COLL}
\cite{DAS JEV}.
\begin{eqnarray}
{\cal L}_{(0)} = - [ \frac{\pi^2}{6} \ \phi^3(x) +
(\frac{U^{\prime^2}(x)}{8} - \mu_F) \phi(x)] +
\frac{1}{4} \int dy \,\frac{U^{\prime}(x)- U^\prime(y)}{x-y} %13
\phi(x)\phi(y)  \nonumber \\
\end{eqnarray}
Here $\mu_F$ is a Lagrange multiplier (the chemical potential) that
enforces the constraint of Eq. (10).
Unlike the case of $d=1$ matter \cite{DAS JEV}, $\mu_F$ is not a
free
parameter whose deviation from a critical value is used to define the
scaling behaviour.  Eq. (11) yields the planar contribution to the genus
expansion.  The term $\frac{1}{8}(U^\prime(x))^2$ is the external
effective potential in which the eigenvalues move.  Its contribution to
the action clearly produces the $\frac{\beta^2}{8}
T_r(U^\prime(\Phi))^2$ term in Eq. (1).  Similarly, the non local
bilinear interaction term in Eq. (11) is the collective field version
for ${\cal H}_{int}$ in Eq. (6).  The cubic term in Eq. (11)
arises
due to the transformation from matrix eigenvalues to collective modes.
If one considers the first few terms in the $1/\beta$ expansion of the
collective field action, including the kinetic term that is identical
to the one used in $d=1$ matrix models, one obtains (up to ambiguities
known in this expansion) a non local collective field theory analogous
to the one developed in \cite{DAS JEV}.

However, unlike the latter, fluctuations of our collective field
theory around the WKB solution do not correspond to genuine string field
components due to the fact that the target space dimension is negative.
Nevertheless, it might include minor fractions of string field
compontents -- namely, discrete states.

The planar collective field equation of motion is readily found to be
\begin{eqnarray}
-\frac{\delta S[\phi]}{\delta\phi(x)}_{\mid_{planar}}=
\frac{\pi^2}{2} \phi(x)^2 + \frac{U^\prime(x)^2}{8} - \mu_F \nonumber \\
%14
-\frac{1}{2} \int dy \frac{U^\prime(x) - U^\prime(y)}{x-y}
\phi(y) = 0
\end{eqnarray}
where $\phi(x)$ is subjected to the constraint of Eq. (10), and that by
definition, $\phi(x)$ is nonnegative.

A crucial observation is that for $\frac{N}{\beta}= 1$ and $\mu_F = 0$,
this nonlinear
nontrivial integral equation is identical to the planar limit of the
Schwinger-Dyson equation obeyed by the loop operator in the original
zero dimensional matrix model \cite{NEUB} \cite{BP IZ} \cite{BE IZ}
\begin{eqnarray}
&&F(z)^2 - U^\prime(z) F(z) + \eta(z) = 0  \nonumber \\
&&F(z) =  \lim_{\beta \approx N \rightarrow \infty} < \frac{1}{\beta}
T_r \frac{1}{z - \Phi} > \ ,  \ \
\eta(z) = \lim_{\beta \approx N \rightarrow \infty} < \frac{1}{\beta}
T_r \frac{U^\prime(z) - U^\prime(\Phi)}{z - \Phi} > \nonumber \\
%15
\end{eqnarray}
when $z$ approaches the real axis. \footnote{
As $z = x - i \epsilon \ , \ \ \epsilon \rightarrow 0+$ we have
\cite{NEUB} \cite{BP IZ} \cite{BE IZ}
\begin{eqnarray}
F(z) = \frac{1}{2} U^\prime(x) + i \pi \phi(x)  \nonumber     %21
\end{eqnarray}
thus making the imaginary part of Eq. (13) vanishing and its real part
proportional to Eq. (12).}

The fact that Eq. (12) is identical to the planar loop equation of the
original matrix model is not surprizing and conforms with the postulates
of \cite{GR HAL}.  Moreover, it seems that the WKB expansion of Eqs. (3)
and (4) should correspond term by term to the genus expansion of the
corresponding Schwinger-Dyson (i.e. loop) equations in the original
model\footnote{It can be shown \cite{JF}, for example, that an Ehrenfest
theorem, associated with Eq. (4), given by $<\frac{\partial}{\partial
\Phi_{cd}} \ \frac{1}{\beta} (\frac{1}{Z-\Phi})_{ab} > \equiv 0$ leads
order
by order in the WKB expansion (in the $N < \beta$ phase) to the genus
expansion of the corresponding loop equation in the original matrix
model, of which Eq. (13) is the leading (planar) term.}

Therefore, under the conditions $\frac{N}{\beta} = 1$ and $\mu_F = 0$,
$\phi(x)$ that solves Eq. (12) is just the planar limit eigenvalue
density of the original Dyson gas in an external potential $U(x)$.

Thus, for a matrix potential $U(\Phi)$ in the universality class of the
k-th multicritical point, $\phi(x)$ will exhibit $k$-th order
multicritical behaviour.  In particular, if we chose
\cite{NEUB}\footnote
{The $(+)$ subscript means that in an expansion of the r.h.s. of Eq.
(14) around $x = \infty$ we keep only nonnegative powers of $x$.}
\begin{eqnarray}
U^\prime_k(x) = \frac{k!(k+1)!}{(2k)!}
[(2 - x)^{k-1}(x^2 - 4)^{\frac{1}{2}}]_+       %14
\end{eqnarray}
then $\phi(x)$ is supported only along the closed segment [-2,2] on the
real axis and is given by
\begin{eqnarray}
\phi_k(x) = \left\{  \matrix{\frac{k!(k+1)!}{2 \pi(2k)!} (4 -     %15
x^2)^{\frac{1}{2}} (2-x)^{k-1} \ &,& \ \ \mid x \mid \leq 2 \cr
0          &,& \mbox{otherwise} } \right.
\end{eqnarray}
and satisfies
\begin{eqnarray}
\int^2_{-2} \phi_k (x) dx = 1 \ .         %16
\end{eqnarray}

Eqs. (14)-(16) are the solution to Eqs. (10) and (12) precisely at the
k-th multicritical point.

In order to have a complete solution of the stabilized model on the
sphere one has to show that these equations are properly deformed by
turning on the cosmological constant.  Namely, that under a deformation
of $Tr U_k(\Phi)$ by the pucture operator $Tr \Phi$, \ \
$Tr U_k(\Phi) \rightarrow TrU_k(\Phi)+ \mu_B Tr \Phi$,
where $\mu_B$ is the bare cosmological constant, there exists a solution
to Eqs. (10) and (12), supported along a single segment on the real
eigenvalue axis, such that $1 - \frac{N}{\beta}$ scales as
$\beta^{-2k/2k+1}$.  Such a deformation of the normalization condition
in Eq. (16) may be obtained by allowing the multicritical end-point
(i.e. $x = 2$) of the support of $\phi_k(x)$ in Eq. (15) to vary, on a
proper scale.\footnote{In the d=1 matrix model \cite{DAS JEV}, variation
of the chemical potential $\mu_F$ changes the location of the classical
turning points which are the end points of supp$\{\phi(x)\}$.}  This
scale
must be that of the double scaled fluctuations of the matrix near its
critical point $\Phi_c = 2$, i.e., \cite{GR MG} \cite{DG SH}
$\beta^{-2/2k+1}$.

Therefore, the required eigenvalue distribution should be supported
along a segment [-2,b] where
\begin{eqnarray}
b = 2 - \epsilon \beta^{-2/2k+1}
\end{eqnarray}
in which $\epsilon$ is a finite real parameter.

Such a deformation of $\phi_k(x)$ alone is not enough to obtain the
desired scaling behaviour of $1 - \frac{N}{\beta}$, since it generically
induces all k-1 relevant deformations \cite{GR MG} \cite{NEUB}  present
at the $k^{th}$ multicritical point.  The desired solution to Eqs. (10)
and (12) must therefore include counterterms that will cancel these
unwanted scaling contributions to $1 - \frac{N}{\beta}$.  Thus, it must
have the general form
\begin{eqnarray}
\phi(x) =
\frac{C_k}{\pi}(2+x)^{\frac{1}{2}}(b-x)^{k-\frac{1}{2}} +     %18
\sum^{k-1}_{n=1} \beta^{\frac{-2(k-n)}{2k+1}} \
\alpha^{(k)}_n \frac{C_n}{\pi} (2+x)^{\frac{1}{2}} (b-x)^{n-\frac{1}{2}}
\end{eqnarray}
on its support [-2,b].

Here
$C_n = \frac{n!(n+1)!}{2(2n)!}$
normalizes $\int^b_{-2} \phi_n(x)$ to unity when $b \rightarrow 2$ as in
Eqs. (15) and (16).

$\beta^{-\frac{2(k-n)}{2k+1}} \ (n \leq k-1)$ are the scaling dimensions
of the relevant perturbations at the k-th multicritical point \cite{GR
MG} \cite{NEUB} ensuring that $\phi(x)$ scales as a whole as
$\beta^{-\frac{2k-1}{2k+1}}$ when $b-x \sim y \beta^{-2/2k+1}$.

Finally, $\alpha^{(k)}_n \ (n \leq k-1)$ are the double scaled couplings
of the relevant scaling operators, that will be uniquely fixed by the
scaling behaviour of $1-\frac{N}{\beta}$.

Indeed, using the elementary integral
$\int^b_{-2} \ \frac{C_n}{\pi} (2 + x)^{\frac{1}{2}} (b-x)^{n -
\frac{1}{2}} dx = (\frac{b+2}{4})^{n+1}$
Eqs. (10), (17) and (18) yield
\begin{eqnarray}
\frac{N}{\beta} = (1- \frac{\epsilon}{4} \lambda)^{k+1} +    %19
\sum^{k-1}_{n=1} \alpha^{(k)}_n \lambda^{k-n} (1 - \frac{\epsilon}{4}
\lambda)^{n+1}
\end{eqnarray}
where we have set $\lambda = \beta^{-\frac{2}{2k+1}}$.

In the vicinity of the k-th multicritical point, $\frac{N}{\beta}$
scales as
\begin{eqnarray}
\frac{N}{\beta} = 1 - t \beta^{-\frac{2k}{2k+1}}          %20
\end{eqnarray}
where $t$ is the renormalized cosmological constant.  Therefore,
expanding the r.h.s. of Eq. (19) in powers of $\lambda$, the
coefficients of $\lambda, \lambda^2, \cdots \lambda^{k-1}$ must vanish,
providing $k-1$ equations that uniquely fix the $k-1$ unknowns
$(\alpha_n)$ in Eq. (18).  The latter are readily  found to be given in
closed form by
\begin{eqnarray}
\alpha^{(k)}_n = \left( \matrix{k+1 \cr n+1}  \right)
(\frac{\epsilon}{4})^{k-n}                              %21
\end{eqnarray}
whence $\frac{N}{\beta} = 1 - (k+1)(\frac{\epsilon}{4})^k \
\beta^{-\frac{2k}{2k+1}} \cdot (1+{\cal O}(\beta^{-\frac{2}{2k+1}}) )$.
Therefore, the renormalized cosmoligical constant is given
by\footnote{Eqs. (21) and (22) were obtained in closed form by M.
Moshe.} \ \ $t=(k+1)(\frac{\epsilon}{4})^k$, or equivalently
\begin{eqnarray}
\epsilon = 4 (\frac{t}{k+1})^{1/k}                      %27
\end{eqnarray}

Eqs. (17), (18), (21) and (22) comprise together the desired deformed
solution $\phi(x)$ to Eq. (12).

To this eigenvalue distribution we may add relevant scaling deformations
with arbitrary couplings, that will not alter the normalization
condition in Eq. (10) \cite{NEUB}.

Eq. (22) exhibits the well known scaling relations \cite{GR MG} on the
sphere between the specific heat $\epsilon$ of the original matrix
model and the cosmoligical constant $t$.  That is, Eq. (22) implies that
the string susceptibility is given by
$\gamma_{str} = - 1/k$.
Equivalently, it implies that it is impossible to obtain a
negative cosmological constant in our solution for even values of $k$,
keeping $\epsilon$ real -- i.e., that it holds only in the $N < \beta$
phase of the theory.

General solutions of the planar loop equation (Eq. (12)) in the $N >
\beta$ phase of multicritical points of even order $k$ would probably
have branch point singularities in the complex eigenvalue plane
(arranged in complex conjugate pairs) other than the two branch points
on the real axis \cite{DA2} \cite{DA3}, giving rise to multicut (planar)
macroscopic loops which may be supported at negative loop lengths where
they are also oscillatory and nonpositive definite).  The latter fact,
should however, in our opinion, be considered as a sickness of the
planar loop equation, or equivalently of the saddle point (WKB) solution
of the stabilized matrix model -- rather than as an {\em a priori}
signal of non-perturbative instabilities of the Marinari-Parisi
procedure.

$\beta\phi(x)$ given by Eqs. (18) and (21) clearly equals the WKB
approximation to the particle density of the interacting Fermi gas of
eigenvalues (integrated up to the Fermi energy).

By construction, $\phi(x)$ scales as $\beta^{-\frac{2k-1}{2k+1}}$ in
terms of the scaling variable $y = \beta^{\frac{2}{2k+1}} (b - x)$.
Therefore, according to well known arguments, the WKB density of
one-particle states at the Fermi energy, corresponding to one-particle
excitations of the ground state of the gas is given by\footnote{Such
one-particle states should exist at least in a frame work of
Hartree-Fock analysis of the gas.}
\begin{eqnarray}
\frac{1}{\beta} (\frac{\partial N}{\partial E})_{\mid_{E=E_F}} \sim
\int^b_{-2} \frac{dx}{\phi(x,b)} \approx \beta^{\frac{2k-3}{2k+1}}
\int^{\infty}_0 \frac{dy}{\phi(y, \epsilon)} =
\beta^{\frac{2k-3}{2k+1}} \rho(\epsilon)           %23
\end{eqnarray}

This integral is generically well behaved, and diverges only as
$\epsilon \rightarrow 0$, i.e., as one approaches the $k$-th
multicritical point.\footnote{$\phi(x)$ given by Eqs. (18) and (21)
obviously vanishes only at $x=b, \ -2$, where it has generically Wigner
semicircle singularities as long as $\epsilon \neq 0$.}

Thus, the energy-gap of these excitations double-scales as
\begin{eqnarray}
\Delta(\epsilon) = [ \frac{1}{\beta}(\frac{\partial N}{\partial
E})_{\mid_{E = E_F}} ]^{-1} = \beta^{-\frac{2k-3}{2k+1}} \
\frac{1}{\rho(\epsilon)} \sim \mid 1 - \frac{N}{\beta}       %30
\mid^{\frac{2k-3}{2k}}
\end{eqnarray}
where the last proportionality is expected on general scaling arguments.
For odd values of $k$, our solution $\phi(x)$ is well defined either in
the $N < \beta$ phase or in the $N > \beta$ phase.  It has the same
singularity structure either for positive $\epsilon$ and $t$ or for
negative
ones.  Thus, in such cases $\Delta(\epsilon)$ in Eq. (24) is well
defined in both phases.

In case of even values of $k$, $\phi(x)$ we have found exists only in
the $N < \beta$ phase where $\Delta(\epsilon)$ is real and positive.
Recall that a positive real double scaled excitation gap is essential
for the stabilization of the model by Eqs. (3) and (4) to work.

For $k=2$ our results reproduce the WKB analysis of \cite{PM(1)}
\cite{KAR MIG} \cite{DA3} in the $N < \beta$ phase of the matrix model.
Indeed, in this case, Eqs. (17), (18), (21) , (22) and (24)  read
\begin{eqnarray}
\phi(x) = \left\{ \matrix{\frac{1}{4\pi} [(2+x)(2-\epsilon\beta^{-2/5}
-x)]^{\frac{1}{2}} (2 + \frac{\epsilon}{2} \beta^{-2/5} -x) && -2 \leq x
\leq 2 - \epsilon \beta^{-2/5}  \cr 0 && \mbox{otherwise} } \right.
%25
\end{eqnarray}
as well as
$\frac{N}{\beta} = 1 -3(\frac{\epsilon}{4})^2 \beta^{-4/5}$
and
$\Delta (\epsilon) = \frac{\sqrt{6 \epsilon}}{4\pi^2} \
\beta^{-1/5} = \frac{1}{\pi^2} (\frac{3}{4} t)^{\frac{1}{4}}
\beta^{-1/5} \sim (1- \frac{N}{\beta})^{\frac{1}{4}}$.

This single segment supported $\phi(x)$ corresponds to the fact that the
Fermi energy of the eigenvalue gas described by Eq. (7), coincides with
the shallow local minimum of $V(x)$ in Eq. (8) near $x = +2$.
This eigenvalue distribution leads to a single cut maxroscopic (planar)
loop operator \cite{NEUB} whose Laplace transform is supported only at
non-negative loop lengths \cite{DA3} \cite{KA}.  Analysis of Eqs. (7)
and
(8) in the $N > \beta$ phase was made in \cite{DA3}.  We have briefly
described it in our comments following Eq. (22), where we have also
stated our interpretation of the difficulties pointed out in
\cite{DA3} \footnote{We have also constructed two solutions to Eq. (12)
for
$k=2$ and $\frac{N}{\beta} > 1$ in which $\mu_F \neq 0$.  One of these
solutions with a single segment support exhibits scaling behaviour of
the $k=3$ point.  The other solution with two real and two complex
conjugate branch points seems to correspond to the one discussed in
\cite{DA3} . }

We have remarked above$^7$ that the  analysis of the $k=2$
point in \cite{KAR MIG} were performed by keeping the original cubic
matrix potential critical while varying $N/\beta$ around its critical
value 1.  At this point of our discussion it is clear that
\cite{KAR MIG} has been successful in doing so because precisely for
that cubic potential does $N/\beta$ couple to its appropriate scaling
perturbation in ${\cal H}_{int}$  -- namely, the puncture
operator $Tr \Phi$.

Up to this point we have established that {\em all stabilized
multicritical
one matrix models are equivalent on the sphere to the corresponding
original (zero dimensional) models}.  This is expected also to hold to
all orders in the genus expansion \cite{GR HAL} \cite{JF} (at least in
the $N < \beta$ phase in the case of even $k$).  This conclusion is a
good starting point and a motivation to study the non-perturbative
nature of the stabilized multicritical models.

It is well known that the set of all multicritical points of (zero
dimensional) one matrix models form an integrable  hierarchy -- namely
that of the KdV equation \cite{GR MG} \cite{DG SH} \cite{DG} \cite{BA DO
SHE}.

The latter may be represented by a set of Virasoro constraints that
annihilate the exact partition function of the matrix model
\cite{LOOPS}.

Thus, in order to establish quantitative results concerning
nonperturbative differences between the original and stabilized matrix
models one  must first try to reformulate the latter as an integrable
hierarchy.

Such a construction, if possible, may be carried either by attempting to
formulate the loop equations of the stabilized models in a manner
analogous to \cite{LOOPS} or by checking whether the interacting Fermi
gas itself forms an integrable system.

In the first case,
one makes an explicit use of the basic observables of the
matrix model -- namely the loop operators -- hence an immediate
comparison of the two types of models may be done.  It may well be that
using the exact collective field theory in this respect, without
expanding its Jacobian in powers of $1/\beta$, turns out to be quite
valuable \cite{JEV 777}, especially due to the fact that this Jacobian
(for the stabilized model) is identical to the one used in d=1 matrix
models \cite{CO DE}.  The enormous symmetry possessed by the collective
field theory in the latter case \cite{COLL W} or by its Fermionic
counterpart \cite{FFLUID} might have counterparts in our case as well
(even though surely in our case we have more complicated Fermi ``Fluid
dynamics'').

The other method proposed above will yield, if it turns out to be
successful, the entire spectrum of the interacting Fermi gas and its
exact S-matrix.  This may be by itself an interesting result in the
theory of $1 + 1$ dimensional integrable models -- extending the local
nonlinear Schr\"{o}dinger model into a nonlocal version (with special
one and two body interactions -- derived from multicritical
potentials).\footnote{Note also that by quantizing the interacting Fermi
gas -- in a Hartree-Fock formalism one obtains, expanding around the WKB
(i.e., planar) Fermi field configuration, a non local version of the the
Thirring model, in an analogous manner to the Dirac theory obtained in
\cite{GR KL1}.} \\

\noindent{\em Summary and Conclusions}\\

We have shown that the Marinari-Parisi definition of pure gravity (k=2)
may be extended to stabilize all higher one matrix multicritical points.
This was done by demonstrating the equivalence on the sphere of the
original and stabilized models.  It seems that this equivalence should
hold to all orders in the genus expansion.  The collective field of the
stabilized matrix model turned out to be useful in coping with
eigenvalue interactions in our semiclassical treatment of the Fermi gas.

We have pointed out that, since the whole structure of multicritical
points of one matrix models may be transferred to the Marinari-Parisi
arena, the most important questions are whether they form there an
integrable hierarchy as the original models do, and if the latter is
answered on the affirmative -- how does it differ from the original KdV
hierarchy?\\

\noindent{\em Acknowledgements}\\

It is a pleasure to thank D. Bar-Moshe, B. Blok, M. Karliner, G. Mandal, M.S.
Marinov, M. Moshe, M. Spiegelglas and S. Yankielowicz for valuable discussions.

I am especially indebted to M. Moshe for his interest and encouragement
during this work.  I also thank the Theory Group at CERN for its
hospitality during the last summer where part of the work was completed.

This work was supported in part by the Fund for Basic Research
administrated by the Israel Academy of Sciences and Humanities.

\end{document}